\documentclass{aa}
\usepackage{graphics}
\begin{document}
\thesaurus{13.18.1; 11.01.2; 11.09.1}
\title{J1343+3758: The third largest FRII-type radio galaxy\\
in the Universe}
 
\author{J. Machalski \and M. Jamrozy}
\offprints{J. Machalski}
\institute{Astronomical Observatory, Jagellonian University,
ul. Orla 171, PL-30244 Cracow, Poland\\
email: machalsk@oa.uj.edu.pl, jamrozy@oa.uj.edu.pl}
\date{Received ....; accepted ...}
\authorrunning{J. Machalski \& M. Jamrozy}
\titlerunning{J1343+3758: The third largest FRII-type radio galaxy}
\maketitle
 
\begin{abstract}
A radio source of the Fanaroff--Riley type II with the angular size of 11.3 arc
min is identified with an optical galaxy at $z=0.229$. Thus, the projected 
linear extent of the radio structure is 3.14 Mpc which makes it the third
largest classical double radio source known after 3C236 and WNB2147+816.
The high-frequency VLA observations, a galaxy identification and its optical
spectroscopy are reported. The equipartition magnetic field and energy density
in the source is calculated and compared with corresponding parameters of other
giants known, indicating that either parameter of the source investigated is 
extremely low. On the other hand, the age estimate of relativistic electrons 
and the advance speed of the lobe material are comparable to the respective 
parameters characterizing other low-luminosity giant sources, as well as much 
smaller and brighter 3CR sources. The environment analysis suggests that 
J1343+3758 lies in a significantly poor region of intergalactic medium.
 
\end{abstract}
 
\section{Introduction}
 
In this letter, we report a discovery of the third largest radio galaxy in the
Universe. Known from years, the record holder radio galaxy 3C236 with the linear
extent of 5.65 Mpc (if $H_{o}=50$ km~s$^{-1}$~Mpc$^{-1}$ and $q_{o}=0.5$ which
we assume hereafter) was mapped with the WSRT over 20 years ago (cf. Willis,
Strom \& Wilson 1974; Strom \& Willis 1980). The second largest, recently
discovered, 3.56-Mpc radio galaxy is WNB2147+816 (Palma et al. 2000). Proceeding
with our investigation of low-luminosity giant radiosource candidates (cf. 
Jamrozy \& Machalski 1999; Machalski \& Jamrozy 2000), selected from the VLA 
surveys: FIRST (Becker, White \& Helfand 1995) and NVSS (Condon et al. 1998), 
we focussed our attention on the source J1343+3758. This source can be discerned 
in the 325-MHz WENSS survey (Rengelink et al. 1997) and 1.4-GHz VLA NVSS survey 
as a 11.3 arc min large FRII-type radio source. The source consists of the two 
extended radio lobes, possibly connected with a very dimmed bridge. At the outer 
edges of both lobes there are more compact bright regions clearly visible on the 
relevant FIRST map J134300+38071E.COADD.1, however from that FIRST map alone
one could not be sure that these bright regions belong to the same source. The
region in the NE lobe contains a very bright hot spot. The source was also
detected during the low-resolution sky surveys at 151 and 232 MHz (6C2: Hales,
Baldwin \& Warner 1988; Zhang et al. 1997, respectively). In the 408 MHz and
4.85 GHz surveys (B3: Ficarra, Grueff \& Tomassetti 1985; GB6: Gregory et al. 1996,
respectively), only its SW lobe was detected above the survey's limit. 

The FIRST map may suggest a radio core with 1.4-GHz flux density of about 1.5 
mJy at the J2000 position: RA $13^{\rm h}42^{\rm m}54\fs 53$, Dec. 
$+37\degr 58\arcmin 18\farcs 1$, but the rms noise on this map (0.15 mJy 
beam$^{-1}$) produced other spots of comparable intensity. Therefore, we have 
made additional high-frequency VLA observations which allow us to confirm 
its radio core and firmly identify the source with a galaxy at the redshift of 
0.229. This VLA observations are described and the radio spectrum of the entire 
source and its lobes are determined in Sect.~2. The optical spectroscopy of the 
identified galaxy is described in Sect.~3. In Sect.~4, the equipartition 
magnetic field and energy density within the lobes as well as the lifetime of 
relativistic electrons at the frequency of 325 MHz and the mean advance velocity 
of the lobes, are calculated and compared with the respective values found for 
other giants as well as for much smaller 3CR sources by Ishwara-Chandra \& 
Saikia (1999). Finally, the environment density is
estimated by the simple counts of galaxies around the identified galaxy and the
calculation of particle density around the radio lobes.
 
\section{Radio observations}

A field centred at the pressumed core position taken from the FIRST map (cf. 
Sect.~1) was observed with the VLA B--array at 4.86 GHz. This observing
frequency and the array configuration did not allow us to map the brightness 
distribution in the extended lobes with their brightest regions lying 5--6 arc 
min apart from the core.
 
\begin{figure}[t]
\resizebox{\hsize}{!}{\includegraphics{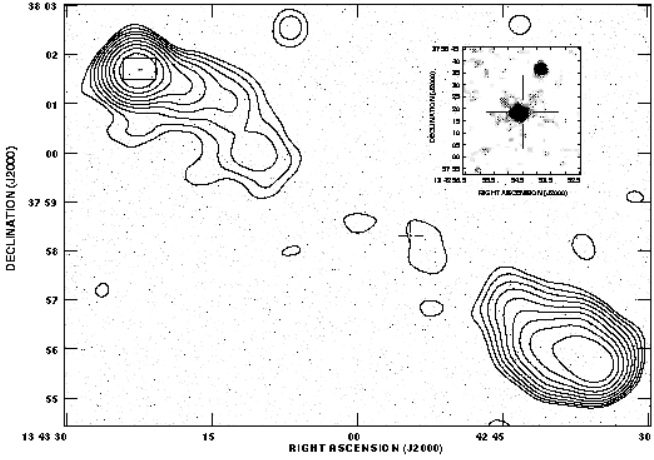}}
\caption{1.4-GHz NVSS contour map. The radio core (marked by the cross) and fine 
structure of the NE hotspot detected at 4.86 GHz are shown in gray scale. The 
small rectangle area marked in the NE lobe is enlarged in Fig.~2. The 
inset shows position of this core in the DSS image of the identified galaxy}
\label{fig1}
\end{figure}

The observations were conducted on December 13, 1999. $\theta_{\rm HPBW}$ of the 
synthesized beam was $1\farcs 24\times 1\farcs 16$. With the
integration time of $3\times 24$ min, the rms fluctuations were about
26 $\mu$Jy~beam$^{-1}$. The $1\farcs 0\times 0\farcs 8$ core component
of 0.7 mJy beam$^{-1}$ and 
 the integral flux density of 1.1 mJy was detected 
at $13^{\rm h}42^{\rm m}54\fs 53$, 
$+37\degr 58\arcmin 18\farcs 8$ (J2000). Besides the core, very much attenuated 
emission was detected from the hotspot in the NE lobe. 

Fig.~1 shows the radio contours of the source taken from the NVSS map 
C1348P36.IQU.1 overlayed on the brightness from our 4.86-GHz map (gray scale).
Because of very large scale of the entire source, poorly visible dot indicating 
the radio core is marked by the cross.  
The most compact structure in the NE lobe, reproduced from the 1.4-GHz FIRST 
map (cf. Sect.~1), and our 4.86-GHz map resolving the hotspot 
are shown in Fig.~2. The deconvolved size of the hotspot at 1.4 GHz is 
$6\farcs 3\times 2\farcs 8$ at PA=+83$\degr$. This hotspot is resolved at 4.86 
GHz into two components with the deconvolved sizes of 
$5\farcs 8\times 2\farcs 7$  and $5\farcs 2\times 3\farcs 1$. These sizes are 
used in Sect.~4 to estimate the area of the bowshock and ambient density of the 
environment in front of a head of the NE lobe.

\begin{figure}[th]
\resizebox{60mm}{!}{\includegraphics{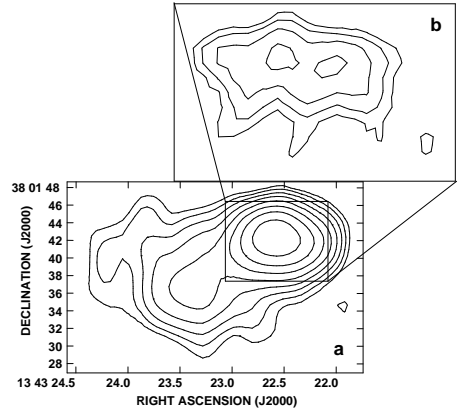}}
\caption{{\bf a} Compact structure in the NE lobe reproduced from the 1.4-GHz 
FIRST map suggesting presence of a hotspot. {\bf b} Our 4.86-GHz map of this 
hotspot}
\label{fig2}
\end{figure}

The degree of asymmetry of J1343+3758 is low.
The ratio of separation between the core and brightest regions in the lobes and 
the misalingnment angle are 1.32 and 3 deg, respectively. The widths of the 
lobes (322 kpc and 274 kpc) are determined using the prescription of Leahy \& 
Williams (1984), i.e. the deconvolved half-power widths are multiplied by the 
factor $2/\sqrt{3}$. The average of the above values give the overall axial 
ratio of 10.5.

The flux densities available for the source at frequencies from 151 MHz to 5 GHz are
given in Table~1.
 
\begin{table}[tbh]
\caption{Radio flux densities of the source and its lobes}
\begin{tabular}{lllll}
\hline
Freq.  & Survey & Total flux  & SW lobe     & NE lobe \\
(GHz]  & Tel.   & [mJy]       & [mJy]       & [mJy]   \\
\hline
0.151  & 6C2    & $630\pm 73$ & $400\pm 64$ & $230\pm 50$ \\
0.232  & Miyun  & $620\pm 120$& $330\pm 100$& $290\pm 90$ \\
0.325  & WENSS  & $389\pm 32$ & $243\pm 20$ & $146\pm 12$ \\
0.408  & B3     &             & $185\pm 35^{a}$ \\
1.4    & NVSS   & $140\pm 9$  & $ 83\pm 4$  & $57\pm 5$ \\
4.85   & GB6    & $36\pm 8$   & $20\pm 4$   \\
\hline
\end{tabular}
Note: a) original B3 flux density is multiplied by 1.087, \\i.e. 
adjusted to the common  scale of Baars et al. (1977)
\end{table}

The integrated spectrum of the entire source and its lobes is shown in Fig.~3.
In order to calculate the total radio luminosity, we fit the observed flux 
density data with an assumed functional form. The best fit to the data in column 
3 of Table 1 is
achieved with a parabola $S(x)$[mJy]$=(-0.199\pm 0.149)x^{2}-(0.809\pm 0.081)x+
(2.256\pm 0.033)$, where $x=\log \nu$[GHz]. This fit gives the best-fitted 1.4
GHz total flux density of 136 mJy, and the fitted spectral indices of $-0.48\pm
0.16$ and $-1.09\pm 0.29$ at 151 MHz and 5 GHz, respectively. Though the quoted 
errors of spectral indices are large, the integrated spectrum clearly steepens 
at high frequencies. A change of the spectrum slope of about 0.5 suggests that 
the break frequency is likely between 151 MHz and 5 GHz, however its too short 
frequency range did not allow a reliable fit of any theoretical spectrum 
accounting for radiative losses to the observations.
 
\begin{figure}
\resizebox{\hsize}{!}{\includegraphics{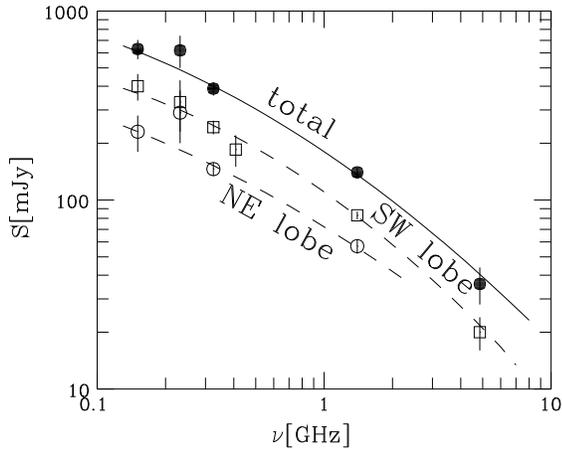}}
\caption{Radio spectrum of the entire source and its lobes}
\label{fig3}
\end{figure}
 
\section{Optical observations}
 
The radio core position coincides perfectly with a galaxy of R=17.94 mag at
$13^{\rm h}42^{\rm m}54\fs 57$, $+37\degr 58\arcmin 19\farcs 2$ (J2000) according to
the magnitude calibration and astrometric position in the Digitized Sky Survey
(DSS) data base. The optical image of this galaxy is shown in the upper-right 
subset of Fig.~1 where the position of the very compact radio core is indicated.

In order to determine a redshift of the identified galaxy, we have made optical
spectroscopic observations with the
2.1m telescope of the McDonald Observatory (Texas). These observations
constituted a part of the larger project involving optical spectroscopy of
the other giant candidates in our sample (cf. Introduction). The `Imaging Grism
Instrument' (IGI) equipped with the TK4 $1024\times 1024$ CCD detector and
cooled with liquid nitrogen was used.
IGI allows direct imaging and spectroscopy with a spatial scale of
$0\farcs 485$ per~pixel within the field of view of 8$\times$ 8 arc min. We
used a grism sensitive to the wavelength range 3750 \AA \,\,to 7600 \AA \,\,and a
2$\arcsec$ wide slit, that provided a dispersion of 3.7 \AA \,\, per~pixel and
spectral resolution of about 12 \AA.
Three exposures of 30 min were taken on two nights of February 28 and 29, 2000,
which allowed to improve S/N ratio without an increase of smearing of the spectrum
during a too long single exposure and due to imperfect tracking of the telescope. The
wavelength calibration was carried out using exposures to helium and mercury
lamps. The flux calibration was provided by exposures of the standard star HZ 44.
The data reduction and calibration were performed with the use of packages of the IRAF
software. The reduced 1-D spectrum is shown in Fig.~4.

\begin{figure}
\resizebox{\hsize}{!}{\includegraphics{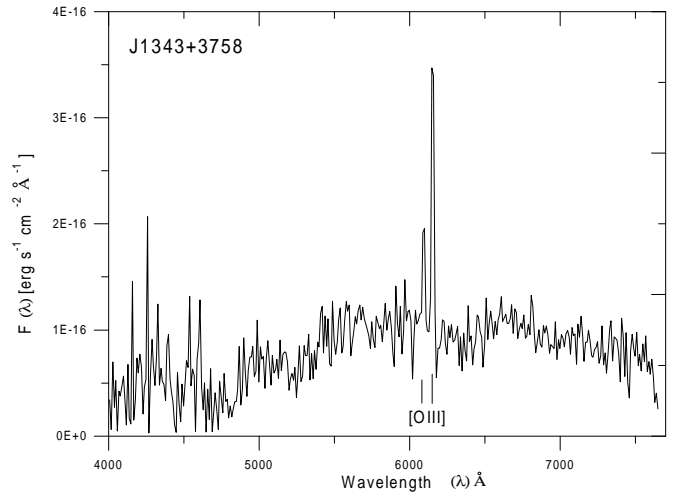}}
\caption{Optical spectrum of the identified galaxy}
\label{fig4}
\end{figure}
 
The spectrum reveals a weak continuum with prominent narrow emission lines [OIII]
4959 \AA \,\,and 5007 \AA, which give a redshift $z=0.229\pm 0.002$.

\section{Physical parameters}
 
Recently, Ishwara-Chandra \& Saikia (1999) have published very interesting 
statistics of some physical parameters calculated for 44 known giants exceeding
1 Mpc, and compared them with the corresponding parameters of smaller 3CR
radio sources. Following Ishwara-Chandra and Saikia, we calculate the
equipartition magnetic field $B_{\rm me}$ and minimum energy density $u_{\rm me}$ in
J1343+3758 using the standard method (e.g. Miley 1980), as well as the ratio
$B_{\rm iC}/B_{\rm me}$, where $B_{\rm iC}=0.324(1+z)^{2}$ nT is the equivalent magnetic
field of the microwave background radiation, and
$B^{2}_{\rm me}/(B^{2}_{\rm iC}+B^{2}_{\rm me})$
which represents the ratio of the energy losses by synchrotron radiation to the
total energy losses due to the synchrotron and inverse Compton processes. 
Following their analysis and assuming: a cylindrical geometry of the source with 
a base diameter of 298 kpc (which is the mean of the deconvolved widths of the 
lobes derived in Sect.~2) and the projected length of 3140 kpc, the filling 
factor of unity, and equal energy distribution between relativistic electrons 
and protons, we found the total radio luminosity of the source between 10 MHz 
and 10 GHz $L_{\rm tot}=2\times 10^{42}$ erg s$^{-1}$ and the source volume of 
$6.4\times 10^{72}$ cm$^{3}$. Since the parabolic fit of the spectrum may 
underestimate fluxes (and thus the source luminosity) at frequencies below 151 
MHz, we also calculated $L_{\rm tot}$ for constant spectral slope of $-0.45$ at 
$\nu <151$ MHz and found only 8 per cent increase of $L_{\rm tot}$. Then we found
$B_{\rm me}=0.083^{+0.028}_{-0.023}$ nT, $u_{\rm me}=(0.63^{+0.43}_{-0.35})\times
10^{-13}$ erg~cm$^{-3}$, $B_{\rm iC}/B_{\rm me}=5.9^{+2.0}_{-1.7}$, and
$B^{2}_{\rm me}/(B^{2}_{\rm iC}+B^{2}_{\rm me})=0.028^{+0.018}_{-0.015}$. The errors in 
$B_{\rm me}$ and $u_{\rm me}$ are calculated adopting errors of 15 per cent in the 
integral luminosity, 0.2 in spectral index, and 50 per cent in the volume.
 
The equipartition magnetic field and energy density found for J1343+3758 are 
extremely low. Only two giants in the sample of Ishwara-Chandra \& Saikia have 
their parameters comparable to the above values. The respective parameters are 
compared in Table 2.

\begin{table}[tbh]
\caption{Comparison of the physical parameters derived for J1343+3758 and NGC6251 
and DA240. The respective parameters of these two latter giants have extremal 
values in the sample of Ishwara-Chandra \& Saikia}
\begin{tabular}{llccc}
\hline
Source  & $B_{\rm me}$  & $u_{\rm me}\,10^{-13}$ & $\frac{B_{\rm iC}}{B_{\rm me}}$ & 
$\frac{B^{2}_{\rm me}}{B^{2}_{\rm iC}+B^{2}_{\rm me}}$ \\
name    & [nT]      & [erg cm$^{-3}$]      \\
\hline
NGC6251 & 0.061     & 0.38               & 5.6      & 0.031 \\
DA240   & 0.077     & 0.61               & 4.5      & 0.046  \\
J1343+3758 & 0.083  & 0.63               & 5.9      & 0.028 \\
\hline
\end{tabular}
\end{table} 

The age of relativistic electrons in the radio lobes $(\tau)$ radiating at a 
given frequency $\nu$ can be deduced from the values of $B_{\rm me}$ and $B_{\rm iC}$. 
Again following the analysis of Ishwara-Chandra \& Saikia, we have calculated  
$\tau$ at $\nu_{o}=325$ MHz. Using their expression (cf. also Alexander \& 
Leahy 1987 and Liu, Pooley \& Riley 1992) which is derived under assumption that 
electrons are isotropized on time-scales much shorter than their radiative 
lifetime (JP model: Jaffe \& Perola 1974), we found $\tau=9.3\times 10^{7}$ yr. 
It is somehow lower than $\langle\tau\rangle\approx 14\times 10^{7}$ yr for 
Ishwara-Chandra \& Saikia's giants with $0.15<z<0.30$, and closer to 
$\langle\tau\rangle\approx 7\times 10^{7}$ yr for those at $z\geq 0.30$. 

However, this is worth to emphasize that the JP model predicts the fastest 
steepening of synchrotron spectrum at high frequencies, while its slowest 
steepening is provided in the case of continuous injection of energetic particles 
(CI model); for the detailed description cf. Myers \& Spangler (1985). Because 
of very limited observational data, one cannot distinguish which model of the 
radiative losses would be plausible for J1343+3758, but this model-dependent 
uncertainty of the synchrotron lifetime of particles in its lobes 
is less than that introduced by other effects like unknown filling factor, 
energy distribution between electrons and protons, or uncertain volume.

If the main axis of the source is close to the plane of the sky, the distance 
from the core to the brightest regions in the lobes will be within 1.4 Mpc and 
1.8 Mpc. Relating $\tau=9.3\times 10^{7}$ yr to any distance between
the above values, the advance speed of the lobes material should be about
$(0.06\pm 0.03)c$. This value is still within the speed range found for much 
smaller, double 3CR radio sources (e.g. Alexander \& Leahy; Liu et al.). 
This advance speed  can suggest that the source achieved its
present size due to expansion in a low-density environment. To check this, we
followed Hill \& Lilly (1991) and estimated the environment density by simple
counts of galaxies around the identified galaxy. We define $N_{1.5}=N_{\rm obs}-
N_{\rm bg}$ as the net excess number of galaxies with $R$ magnitude from
$R_{\rm id}$ to $R_{\rm id}+3$ and within 1.5 Mpc radius around our galaxy. Using the
DSS data base and adopting $R_{\rm id}=17.94$ mag, we found 30 galaxies to meet
the above criterium. However, the number of galaxies with $R>20$ mag can be
underestimated in the DSS as they can be seen at the POSS E--plates only,
therefore we adopt $N_{\rm obs}=30^{+20}_{-8}$. The relevant number of
background galaxies, $N_{\rm bg}=59\pm 12$ was found using the differential counts
$\log [N(R)=0.39R-4.8$ deg$^{-2}$mag$^{-1}$] (Tyson 1988). Resultant negative
net value
$N_{1.5}=-29^{+23}_{-14}$ indicates that J1343+3758 lies in a distinctly poor
region of intergalactic medium. This conclusion is further supported by an estimate
of the particle density around the lobes. Following Lacy et al. (1993), we
assume that the heads of lobes are ram-pressure confined. Taking the deconvolved
diameter of the hotspot in the NE lobe as 3 arc sec, we can estimate the ambient
density $\rho_{\rm a}\stackrel{>}{\sim}1.4~10^{-31}$ g~cm$^{-3}$. This estimate is lower by
an order than that found for the other giants (cf. Parma et al. 1996: Mack et al.
1998; Schoenmakers et al. 1998).
 
\begin{acknowledgements}
Authors acknowledge (i) the National Radio Astronomy Observatory (Socorro, NM)
for the target-of-opportunity observing time, (ii) the National Optical Astronomy
Observatories (Kitt Peak, AZ) for the usage of the IRAF software, (iii) the
Space Telescope Science Institute for the usage of the DSS data base, and (iv) 
Dr U. Klein for his constructive remarks and suggestions improving this paper. 
This work was
supported in part by the State Committee for Scientific Research (KBN) under the
contract PB 0266/PO3/99/17.
\end{acknowledgements}


\begin{thebibliography}{}
 
\bibitem[1987]{alexander}
Alexander P., Leahy J.P., 1987, MNRAS 225, 1
 
\bibitem[1977]{baars}
Baars J.W.M., Genzel R., Pauliny-Toth I.I.K., Witzel A., 1977, A\&A 61, 99
 
\bibitem[1995]{becker}
Becker R., White R., Helfand D., 1995, ApJ 450, 559
 
\bibitem[1998]{condon}
Condon J.J., Cotton W.D., Greisen E.W., et al., 1998, AJ 115, 1693
 
\bibitem[1985]{ficarra}
Ficarra A., Grueff G., Tomassetti G., 1985, A\&AS 59, 255
 
\bibitem[1996]{gregory}
Gregory P.C., Scott W.K., Douglas K., Condon J.J., 1996, ApJS 103, 427
 
\bibitem[1988]{hales}
Hales S.E.G., Baldwin J.E., Warner P.J., 1988, MNRAS 234, 919
 
\bibitem[1991]{hill}
Hill G.J., Lilly S.J., 1991, ApJ 367, 1
 
\bibitem[1999]{ishwara}
Ishwara-Chandra C.H., Saikia D.J., 1999, MNRAS 309, 100
 
\bibitem[1973]{jaffe}
Jaffe W.J., Perola G.C., 1974, A\&A 26, 423

\bibitem[1999]{jamrozy}
Jamrozy M., Machalski J., 1999, Acta Astron. 49, 181
 
\bibitem[1993]{lacy}
Lacy M., Rawlings S., Saunders R., Warner P.J., 1993, MNRAS 264, 721
 
\bibitem[1984]{leahy}
Leahy J.P., Williams A.G., 1984, MNRAS 210, 929

\bibitem[1992]{liu}
Liu R., Pooley G., Riley J., 1992, MNRAS 257, 545
 
\bibitem[2000]{machalski}
Machalski J., Jamrozy M., 2000, In: The Universe at Low Radio Frequencies
(Proc. IAU Symp. No.199, Pune, India), ed. ASP Conference Series (in print)
 
bibitem[1998]{mack}
Mack K.H., Klein U., O'Dea C.P., Willis A.G., Saripalli L., 1998, A\&A 329, 431
 
\bibitem[1980]{miley}
Miley G.K., 1980, ARA\&A 18, 165

\bibitem[1985]{myers}
Myers S.T., Spangler S.R., 1985, ApJ 291, 52
 
\bibitem[2000]{palma}
Palma C., Bauer F.E., Cotton W.D., et al., 2000, prep. astro-ph/0002033
 
\bibitem[1997]{rengelink}
Rengelink R., Tang Y., de Bruyn A.G., et al., 1997, A\&AS 124, 259
 
\bibitem[1998]{schoen}
Schoenmakers A.P., Mack K.H., Lara L., et al., 1998, A\&A 336, 455

\bibitem[1980]{strom}
Strom R.G., Willis A.G., A\&A 85, 36
 
\bibitem[1988]{tyson}
Tyson J.A., 1988, AJ 96, 1

\bibitem[1974]{willis}
Willis A.G., Strom R.G., Wilson A.S., 1974, Nature 250, 625
 
\bibitem[1997]{zhang}
Zhang X., Zheng Y., Chen H., et al., 1997, A\&AS 121, 59
 
\end{thebibliography}
\end{document}